\documentclass[prd,preprint,aps,amsfonts,amssymb,nofootinbib,tightenlines]{revtex4}

\usepackage{amscd}
\usepackage{amsmath}
\usepackage{bm}
\usepackage[dvips]{color}                 
\usepackage{graphicx}
\usepackage{eufrak}
\usepackage{eucal}

\newcommand{\beq}{\begin{equation}}
\newcommand{\eeq}{\end{equation}}
\newcommand{\bea}{\begin{eqnarray}}
\newcommand{\eea}{\end{eqnarray}}

\newcommand{\nn}{\nonumber}
\newcommand{\benn}{\begin{displaymath}}
\newcommand{\eenn}{\end{displaymath}}
\newcommand{\ket}[1]{| #1 \rangle}                     

\def\[{\left[}
\def\]{\right]}

\begin{document}
\preprint{LBNL-54231}

\title{Cold asymmetrical fermion superfluids}

\author{Heron Caldas\footnote{Email: {\tt hcaldas@ufsj.edu.br}.
Permanent address: Universidade Federal de Sao Joao del Rey, S\~{a}o Jo\~{a}o del Rei, 36300-000, MG, Brazil.} }
\affiliation{Lawrence-Berkeley Laboratory, Berkeley, CA 94720, U.S.A.}

\begin{abstract}
In this work we investigate the general properties and the ground state of an asymmetrical dilute gas of cold fermionic atoms, formed by two particle species having different densities. We have shown in a recent paper, that a mixed phase composed of normal and superfluid components is the energetically favored ground state of such a cold fermionic system. Here we extend the analysis and verify that in fact, the mixed phase is the preferred ground state of an asymmetrical superfluid in various situations. We predict that the mixed phase can serve as a way of detecting superfluidity and estimating the magnitude of the gap parameter in asymmetrical fermionic systems.
\end{abstract}
\maketitle
\bigskip

\section{Introduction}

The recent experimental advances in cold atomic traps have induced a great amount of interest in fields from condensed matter to particle physics, including approaches and prospects from the theoretical point of view. The advent of new techniques in dealing with cold Fermi atoms in traps in the last few years \cite{Marco,Truscott,Schreck,Granade,Hadzibabic,Roati,Thomas} have revived the interest in some fundamental questions in many-body quantum physics. The ability of preparing a trap with a strongly interacting ultra-cold gas of fermionic atoms \cite{Thomas} constitutes an example of such an experimental development. In this scenario one may found the necessary conditions for pair formation between species whose Fermi surfaces do not match. An immediate question, which arises in this context, is {\it what is the ground state of a fermion system composed by two components having different masses and densities (or chemical potentials)?}

In this paper we study the possibility and the prerequisites for pair formation between two species of particles with unequal number densities or chemical potentials. These different species could be two fermionic atoms ($^{40}{\rm K}$ or $ ^{6}{\rm Li}$) \cite{Marco,Thomas,Granade} or hyperfine states of the same atom \cite{pethick_book}. The nucleus of neutron stars could also have the basic ingredients for asymmetric pairing, since quarks with different flavors would have different densities \cite{krishna_review,Rischke:2003mt,Schafer:2003vz}.

The pairing between particles with opposite spin and equal and opposite momenta near their (common) Fermi surface, is described by the standard
Bardeen-Cooper-Shrieffer (BCS) \cite{BCS} theory of superconductivity. A model exhibiting asymmetry between the (fixed) chemical potentials of two particle species was first studied by Sarma a long time ago at zero and finite temperature \cite{Sarma}. The Sarma phase was proposed recently as a candidate for the ground state of a system of two species of cold fermionic atoms \cite{Liu1,Liu2}. This state of matter, named ``internal gap'' phase, would be composed by a (homogeneous) mixture of normal and superfluid phases.

It has been showed in a recent paper \cite{PRL1} that the ground state of such an asymmetrical fermionic system is: {\bf I.} either the normal or the BCS states, depending on the asymmetry between the Fermi surfaces of the two species, for the situation of fixed chemical potentials, {\bf II.} a (inhomogeneous) mixed phase formed by normal and superfluid components, for the case of fixed number of particles and finally {\bf III.} a mixed phase having as limits the BCS state for $m_a=m_b$ and normal phase with only $b$ particles when $m_b \gg m_a$, for fixed overall number density.

It has been shown \cite{Liu1} that the coupling constant in the internal gap (Sarma phase) has a lower value, $g_{min}$, instead of being arbitrarily small as in the BCS phase. We show in this paper that the coupling constant supported by the Sarma phase also has a maximum value, $g_{max}$. Thus, the Sarma phase can exist only in the window $g_{min}<g<g_{max}$. As a consequence, the Sarma phase is not favorable for weak enough attractive interactions. For the sake of comparison, we show that the approximate number of particles available to form pairs in the Sarma phase is always less than the one in the BCS phase. As a result, the Sarma phase may have a disadvantage in the condensation energy, since the energy of the system is lowered with pairing. Even for the case of fixed chemical potentials, we found a particular combination for the Fermi surfaces $P_F^a$ and $P_F^b$ of the two species, for which there exist a mixture of normal and BCS phases. In the case of fixed number of particles, we found that the state of lower energy satisfies the conditions for equilibrium between the two components of the mixed phase.

We notice that another possibility for the ground state in asymmetric matter is the Larkin, Ovchinnikov, Fulde and Ferrel (LOFF)-phase~\cite{larkin}. The gap parameter in the LOFF phase also breaks translational invariance, as the mixed phase does. However, since the LOFF phase can exist only in a very narrow window of asymmetry for the chemical potentials we do not consider this possibility in our discussion.

The counterpart for ordering in quantum chromodynamics (QCD) (for the case where the quarks have approximately the same Fermi surface) at high baryon
density is the color superconductivity \cite{krishna_review,Alford:2001dt,Schafer:2003vz,Rischke:2003mt}. For high density asymmetric quark matter the
analog of LOFF state leads to crystalline color superconductivity \cite{QCD-2}. In QCD with two light flavors, the ground state is the two-flavor color
superconductor (2SC) \cite{Alford}.

Still in quark matter, it has been argued that the gapless color superconductivity (analogous to the Sarma phase) can reach stability, under the conditions of color and electrical charge neutrality, in the two~\cite{Igor1,Igor2} and three~\cite{Mark} flavor cases.

The paper is organized as follows. In Section II we introduce the model Hamiltonian and derive the free energy and gap equation by variational and
mean field approximations. In Section III we search for the fundamental state of the asymmetrical fermion superfluid in three situations: fixed chemical
potentials, fixed particle numbers and fixed overall density. We discuss the results and conclude in Section IV.

\section{The Model, basic definitions and the gap equations}

To begin with, let us consider a nonrelativistic dilute (i.e., the particles interact through a short-range attractive interaction) cold fermion system,
described by the following Hamiltonian
\begin{equation}
\label{eq-1}
{\cal H}=H-\sum_{k,\alpha}\mu_{\alpha}n_{\alpha}=\sum_{k} {\varepsilon}^{a}_k a^{\dagger}_{k} a_{k}+{\varepsilon}^{b}_k b^{\dagger}_{k} b_{k} + g \sum_{k,k'} a^{\dagger}_{k'} b^{\dagger}_{-k'} 
b_{-k} a_{k},\,
\end{equation}
where $a^{\dagger}_{k}$, $a_k$ are the creation and annihilation operators for the $a$ particles (and the same for the $b$ particles)
and ${\varepsilon}^{\alpha}_k$ are their dispersion relation, defined by ${\varepsilon}^{\alpha}_k=\xi_k^{\alpha}-\mu_{\alpha}$,
with $\xi_k^{\alpha}=\frac{k^2}{2m_{\alpha}}$ and $\mu_{\alpha}$ being the chemical potential of the $\alpha$-specie, $\alpha=a,b$.
To reflect an attractive (s-wave) interaction between particles $a$ and $b$ we take $g<0$. We shall derive below the state encountered by
an asymmetric fermion superfluid at zero temperature, i.e., when its energy is a minimum. This investigation shall be done in two scenarios
representing different physical situations, namely fixed chemical potentials and fixed number of particles.

\subsection{Solution by a variational procedure}

The BCS ground state, which describes a superposition of empty and occupied (paired) states, is given by
\begin{equation}
\label{bcs}
\left| BCS \right\rangle=\prod_{k}\left[u_k+v_k a^{\dagger}_{k} b^{\dagger}_{-k}\right]\left|0\right\rangle ,
\end{equation}
where the arbitrary complex (a priori) coefficients $u_k$ and $v_k$ are to be determined by a variational calculation. They are subjected to the
normalization, $|u_k|^2+|v_k|^2=1$, and spin singlet, $u_k=u_{-k}$, $v_k=v_{-k}$, conditions.
These constraint are satisfied if one sets $u_k=\sin \theta_k$ and $v_k=\cos \theta_k$ \cite{Tinkham}.

The free-energy (or thermodynamical potential) for this system is obtained as usual by taking the expectation value of ${\cal H}$

\begin{equation}
\label{bcs1}
{\rm F} \equiv <{\rm BCS}|{\cal H}|{\rm BCS}>=\sum_{k,\alpha}{\varepsilon}^{\alpha}_k  \cos^2\theta_k +
\frac{g}{4}\sum_{k,k'} \sin2\theta_k \sin2\theta_{k'}.
\end{equation}
When the minimum of ${\rm F}$ with respect to $\theta_k$ is required ($\frac{\partial {\rm F}}{\partial \theta_k}=0$), we get $\tan 2\theta_k =
g \frac {\sum_{k'} \sin2\theta_{k'}}{2 \varepsilon_k^+}$, where $\varepsilon_k^+=\frac{\varepsilon_k^a+\varepsilon_k^b}{2}=
\frac{1}{2}\left(\frac{k^2}{2M}-\mu \right)$, with $M=\frac{m_a m_b}{m_a+m_b}$ being the reduced mass and $\mu=\mu_a+\mu_b$.
Defining the gap parameter as
\begin{equation}
\label{bcs2}
\Delta= -\frac{g}{2} \sum_{k'} \sin2\theta_{k'} ,
\end{equation}
it follows that $\tan 2\theta_k = - \frac{\Delta}{\varepsilon_k^+}$. This last equation combined with the normalization condition gives

\begin{equation}
\label{bcs3}
u_k^2=\frac{1}{2} \left(1 + \frac{\varepsilon_k^+}{E} \right) = 1 - v_k^2,
\end{equation}
where $E=\sqrt{{\varepsilon_k^+}^2+\Delta^2}$. Substituting these quantities back in Eq. (\ref{bcs1}) we obtain the final expression for the free-energy

\begin{equation}
\label{eq3}
{\rm F}=\int \frac{d^3 k}{(2 \pi)^3} \varepsilon_k^+ \left(1-\frac{\varepsilon_k^+}{E} \right) + \frac{\Delta^2}{g},
\end{equation}
together with the gap equation

\begin{equation}
\label{eq4}
1=-\frac{g}{2} \int \frac{d^3 k}{(2 \pi)^3} \frac{1}{\sqrt{{\varepsilon_k^+}^2+\Delta^2}}.
\end{equation}
Following Ref.\cite{Pape}, the gap equation can be related to the two-body scattering length $a$, in dimensional regularization technique, and we have

\begin{equation}
\label{eq5}
\frac{M}{2 \pi a}=-\frac{1}{2} \int \frac{d^3 k}{(2 \pi)^3} \frac{1}{\sqrt{{\varepsilon_k^+}^2+\Delta^2}}.
\end{equation}
The change of variable $\frac{k^2}{2M\mu}=z$ allows us to write $\int \frac{d^3 k}{(2 \pi)^3}
\to \frac{M \mu k_F}{2 \pi^2}\int_0^{\infty}z^{\frac{1}{2}}dz$, where $k_F \equiv \sqrt{2M\mu}$, so that

\begin{equation}
\label{eq6}
\frac{\pi}{k_F a}=- f_{1/2} \left(\frac{2 \Delta}{\mu} \right),
\end{equation}
where
\begin{equation}
\label{eq7}
f_{\beta}(x)=\int_0^{\infty} dz \frac{z^{\beta}}{\sqrt{(z-1)^2+x^2}}=-\frac{\pi}{\sin \beta \pi}(1+x^2)^{\beta /2} P_{\beta} \left( -\frac{1}{\sqrt{1+x^2}}\right),
\end{equation}
with $P_{\beta}$ being the Legendre function. In the weak coupling limit, $\Delta / \mu << 1$, we expand the function above and obtain, to leading order,

\begin{equation}
\label{eq8}
\Delta \approx \frac{4 \mu}{e^2} e^{\frac{-\pi}{2 k_F |a|}}.
\end{equation}
The free-energy, Eq. (\ref{eq3}), can be obtained in the same way by the use of Eq. (\ref{eq7}):
\begin{align}
\label{eq9}
{\rm F}&=\frac{\Delta^2}{g} - \frac{M \mu^2 k_F}{4 \pi^2} \times \left[f_{5/2} \left(\frac{2 \Delta}{\mu} \right)-2 f_{3/2} \left(\frac{2 \Delta}{\mu} \right)
+f_{1/2} \left(\frac{2 \Delta}{\mu} \right) \right]\\
\nonumber
       &\approx -\frac{k_F^5}{30 \pi^2 M}-\frac{M k_F }{2 \pi^2} \Delta^2.
\end{align}
For the number densities we have
\begin{align}
\label{eq10}
n_{\alpha}&=-\frac{\partial {\rm F}}{\partial \mu_{\alpha}}=-\frac{\partial {\rm F}}{\partial \mu}
\frac{\partial {\mu}}{\partial \mu_{\alpha}}=-\frac{\partial {\rm F}}{\partial \mu}\\
\nonumber
&=\frac{k_F^3}{6 \pi^2}+\frac{M^2 \Delta^2}{2 \pi^2 k_F} \left(5 + \frac{\pi}{ |a| k_F} \right) = n.
\end{align}
Then the particles $a$ and $b$ have the same density in the BCS phase, as it should be. The energy of the superfluid state is

\begin{align}
\label{eq11}
{\rm E}&={\rm F} + \sum_{\alpha} \mu_{\alpha} n_{\alpha}= {\rm F} +  \mu n\\
\nonumber
&=\frac{k_F^5}{20 \pi^2 M}+\frac{k_F^3 \Delta^2}{8 \pi^2 \mu} \left(3 + \frac{\pi}{ |a| k_F} \right).
\end{align}
We can invert Eq. (\ref{eq10}) expressing $\mu =f(n)$ so we can write the energy as a function of the number density

\begin{align}
\label{eq12}
{\rm E}(n)&=\frac{(6 \pi^2 n)^{5/3}}{20 \pi^2 M}-\frac{M}{2 \pi^2} (6 \pi^2 n)^{1/3} \Delta^2\\
\nonumber
&=\frac{(6 \pi^2 n)^{5/3}}{20 \pi^2 M} \left[1-40 e^{\frac{- \pi}{|a| (6 \pi^2 n)^{1/3}}-4} \right] \equiv {\rm E}^{\rm BCS}(n).
\end{align}
Taking the difference in energy between the superconducting and the normal state we find the condensation energy
\begin{equation}
\label{cond}
 {\rm E}^{\rm BCS}(n)- {\rm E}^{\rm N}(n)=-\frac{M}{2 \pi^2} (6 \pi^2 n)^{1/3} \Delta^2.
\end{equation}

\subsection{Solution by the (Bogoliubov) canonical transformation}

Since we are considering a dilute system, we can make the mean-field approximation \cite{Tinkham,Wu}. Defining $\Delta=-g \sum_k <b_{-k} a_{k}>$ that
is chosen as being real, we get
\begin{equation}
\label{eq15}
{\cal H}=\sum_{k} {\varepsilon}^{a}_k a^{\dagger}_{k} a_{k}+{\varepsilon}^{b}_k b^{\dagger}_{k} b_{k}
-\Delta (a^{\dagger}_{k} b^{\dagger}_{-k} + b_{-k} a_{k})-\frac{\Delta^2}{g}.
\end{equation}
Since now ${\cal H}$ is quadratic, it can be diagonalized

\begin{equation}
\label{eq16}
{\cal H}=\sum_{k} {\cal E}^{\alpha}_k {\alpha}^{\dagger}_{k} {\alpha}_{k} + {\cal E}^{\beta}_k {\beta}^{\dagger}_{k} {\beta}_{k} +
\varepsilon_k^b -{\cal E}^{\beta}_k - \frac{\Delta^2}{g},
\end{equation}
where ${\cal E}_k^{\alpha,\beta}$ are the quasi-particle excitations (QPE)
\begin{equation}
\label{eq17}
{\cal E}_k^{\alpha,\beta}=\pm \varepsilon_k^- + E =\pm \varepsilon_k^- + \sqrt{{\varepsilon_k^+}^2+\Delta^2},
\end{equation}
where $\varepsilon_k^-=\frac{\varepsilon_k^a-\varepsilon_k^b}{2}$.
The (linear) transformation used in the diagonalization of ${\cal H}$ is defined as
\beq
\begin{pmatrix}
\alpha_{k} \\
\beta^\dagger_{-k}
\end{pmatrix}
=
\begin{pmatrix}
u_k & -v_k\\
v_k & u_k
\end{pmatrix}
 \begin{pmatrix}
a_{k} \\
b^\dagger_{-k}
\end{pmatrix}
, \eeq
with $u_k$ and $v_k$ given by Eq. (\ref{bcs3}).
\begin{figure}[t]
\includegraphics[height=2in]{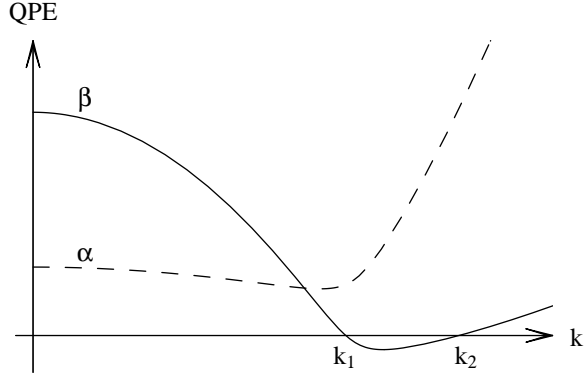}
\caption{\label{fig1}\textit{  Dispersion relation for the quasi-particles
$\alpha$ and $\beta$ showing a region where ${\cal E}^\beta_k$ is negative for
$m_b= 7 m_a$, $P_F^b=1.45\, P_F^a$, and $\Delta=0.29\Delta_0$, obtained by Eq. (\ref{eq26}). Solid curve correspond to ${\cal E}_k^\beta$
and dashed curve correspond to ${\cal E}_k^\alpha$.} }
\end{figure}
Depending on the relative magnitudes of $P_F^b$, $P_F^a$, $m_b$ and $m_a$, either ${\cal E}_k^{\alpha,\beta}$ have two roots,
where $P_F^{\alpha}=(2 m_{\alpha} \mu_{\alpha})^{\frac{1}{2}}$ are the particles Fermi surfaces.
We chose $P_F^b>P_F^a$ and $m_b \geq m_a$ such that only ${\cal E}^{\beta}_k$ crosses zero. The roots of ${\cal E}^{\beta}_k$
can be determined from the equation
\begin{equation}
\label{eq18}
\varepsilon_k^a \varepsilon_k^b=-\Delta^2.
\end{equation}
They are define as $k_1$ and $k_2$,
\begin{eqnarray}
\label{eq19}
k_{1,2}^{2}=\frac{1}{2} ({P_F^a}^2+{P_F^b}^2) \mp \frac{1}{2} [({P_F^b}^2-{P_F^a}^2)^2-16 m_a m_b \Delta^2]^{1/2}.
\end{eqnarray}
Fig.~\ref{fig1} shows that, for some values of
$\Delta$, ${\cal E}^\beta_k$ may be negative for momenta $k_1\le k\le k_2$. A simple inspection in $k_{1,2}$ shows that $\Delta$ has a
critical (maximum) value, above which $k_{1,2}$ are not real,
\begin{equation}
\label{eq20}
\Delta_c=\frac{|\delta p_F|}{4\sqrt{m_a m_b}},
\end{equation}
where $\delta p_F\equiv {P_F^b}^2-{P_F^a}^2$.

The consequence of this is that we can have two physically distinct situations:
\subsubsection{Symmetrical superfluid.} The first situation is when $\Delta \geq \Delta_c$.
In this case ${\cal E}_k^{\alpha,\beta}$ are both positive and we have the standart BCS with Eq. (\ref{eq16}) reduced to
\begin{equation}
\label{eq20-1}
{\cal H} \to {\cal H}^{BCS}=\sum_{k} \varepsilon_k^+ - E - \frac{\Delta^2}{g},
\end{equation}
with the gap parameter given by Eq.(\ref{eq8}). The ``average" Fermi surface for the asymmetrical BCS superfluid can be rewritten as
\begin{equation}
\label{pf}
k_F=\sqrt{{P_F^a}^2+\frac{m_a}{m_a+m_b}\delta p_F}.
\end{equation}
Although we could write the gap with $k_F$ depending on $\delta p_F$ by the equation above, it is a not increasing function
of $\delta p_F$ (with ${P_F^a}$, $m_a$ and $m_b$ kept fixed). As it will become clear below, the BCS free-energy minimum does not depend
on the asymmetry between the Fermi surfaces of the two species until a point where there is a first order phase transition to the normal phase.

\subsubsection{Asymmetrical superfluid.} The second situation, that we call ``Sarma phase" (first pointed out in Ref.~\cite{Sarma}), happens when $\Delta < \Delta_c$. In this case ${\cal E}_k^{\beta}<0$ between $k_{1,2}$.
The free energy in the Sarma phase is obtained when we find a state $\left| \Psi\right\rangle$ which minimizes the diagonalized Hamiltonian,
Eq. (\ref{eq16}). The smallest energy is reached when the modes with negative ${\cal E}^{\alpha,\beta}_k$ are filled and the remaining modes are
left empty. More precisely, the ground state $\ket{\Psi}$ satisfies
\begin{align}
\alpha_k,\beta_k \ket{\Psi}  &= 0 \quad \text{if}
\quad {\cal E}^{\alpha,\beta}_k >0,\nn\\
\alpha_k^\dagger,\beta_k^\dagger \ket{\Psi} &= 0 \quad \text{if}
\quad {\cal E}^{\alpha,\beta}_k <0.
\end{align}
This state can be written in terms of the original $a^{\dagger}_{k}$ and $b^{\dagger}_{k}$ operators and the vacuum state $\ket{0}$ as
\begin{equation}
\label{statesarma}
\left| \Psi \right\rangle=\prod_{\substack{k<k_1 \\ k>k_2}}\left[u_k+v_k a^{\dagger}_{k} b^{\dagger}_{-k}\right] \prod_{k_1}^{k_2}b^{\dagger}_{k}\left|0\right\rangle .
\end{equation}
The state above corresponds to having BCS pairing in the modes $k$ where ${\cal E}_k^{\alpha,\beta}>0$ and a state filled with particles $b$ (a)
in the modes where ${\cal E}_k^{\beta}<0$ (${\cal E}_k^{\alpha}<0$). The free-energy of the Sarma phase turns out to be
\begin{equation}
\label{eq21}
\langle \Psi | {\cal H} \left| \Psi \right\rangle = {\rm F}^S=\int_{\substack{k<k_1 \\ k>k_2}} \frac{d^3k}{(2 \pi)^3} ({\varepsilon}_k^+ - E)+
\int_{k_1}^{k_2} \frac{d^3k}{(2 \pi)^3} {\varepsilon}_k^b -\frac{\Delta^2}{g} ,
\end{equation}
with $\frac{\partial {\rm F}^S}{\partial \Delta}=0$ (remembering that the partial derivative also hits $k_1$ and $k_2$ in the limits of the integrals)
we obtain the gap equation
\begin{equation}
\label{eq22}
1=-\frac{g}{2} \int_{\substack{k<k_1 \\ k>k_2}} \frac{d^3 k}{(2 \pi)^3} \frac{1}{\sqrt{{\varepsilon_k^+}^2+\Delta^2}}.
\end{equation}
Again, the standard BCS result can be recovered for $\Delta \geq \Delta_c$, in which case the equation above reduces to Eq. (\ref{eq4}) and the free
energy goes to
\begin{equation}
\label{eq21-1}
{\rm F}^{S} \to {\rm F}^{BCS}=\int \frac{d^3k}{(2 \pi)^3} ({\varepsilon}_k^+ - E) -\frac{\Delta^2}{g} .
\end{equation}
The equation above can be written in the form of Eq. (\ref{eq3}) using the gap equation (\ref{eq4}), since Eq. (\ref{eq3}) is valid only at the minimum,
due to the variational method we employed in the previous subsection.

\subsubsection{Solving the gap equation}
\begin{figure}[t]
\includegraphics[height=2.in]{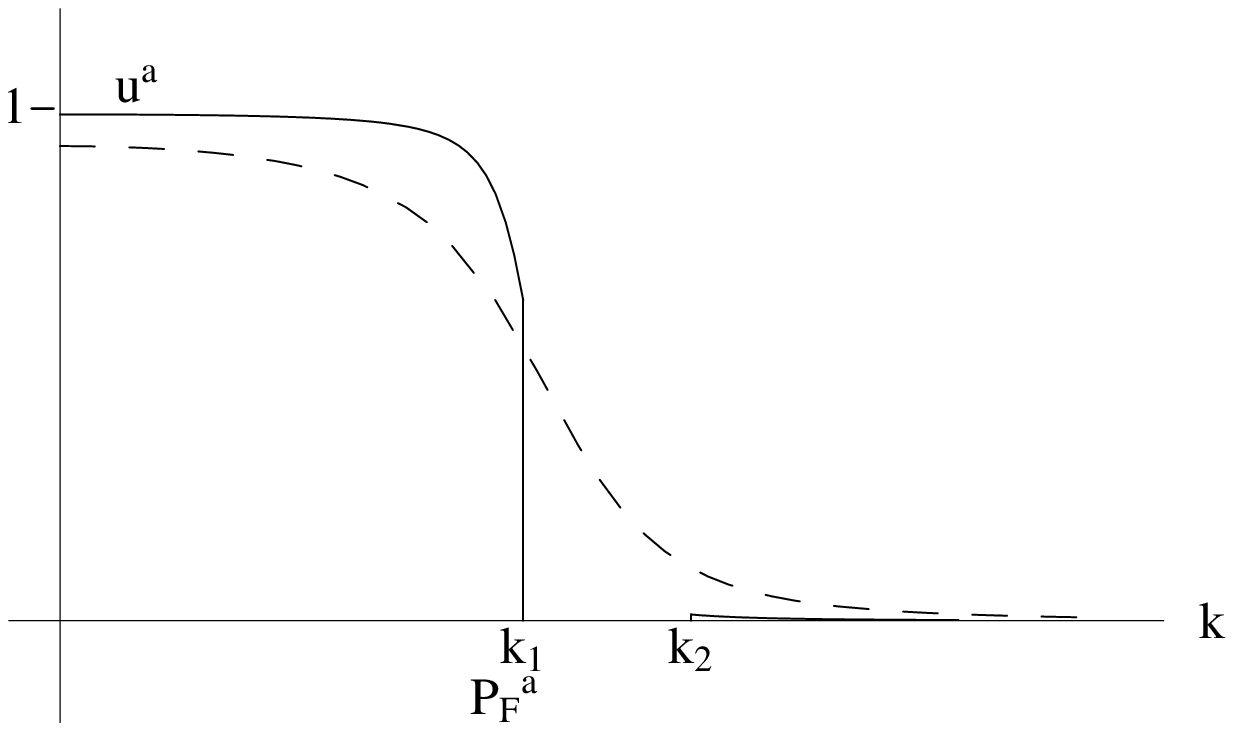}
\hspace{0.2in}
\includegraphics[height=1.5in]{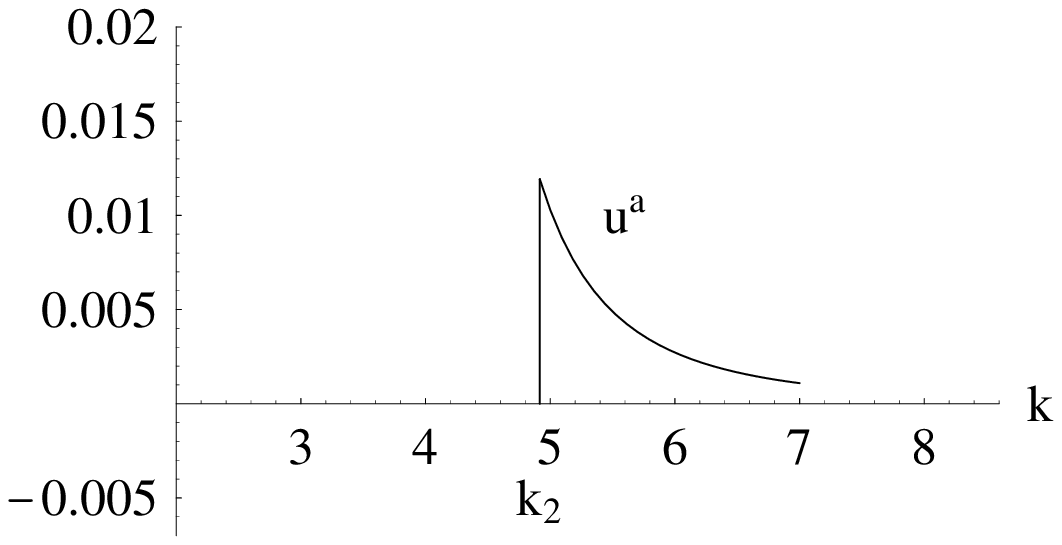}
\hspace{0.2in}
\includegraphics[height=2.in]{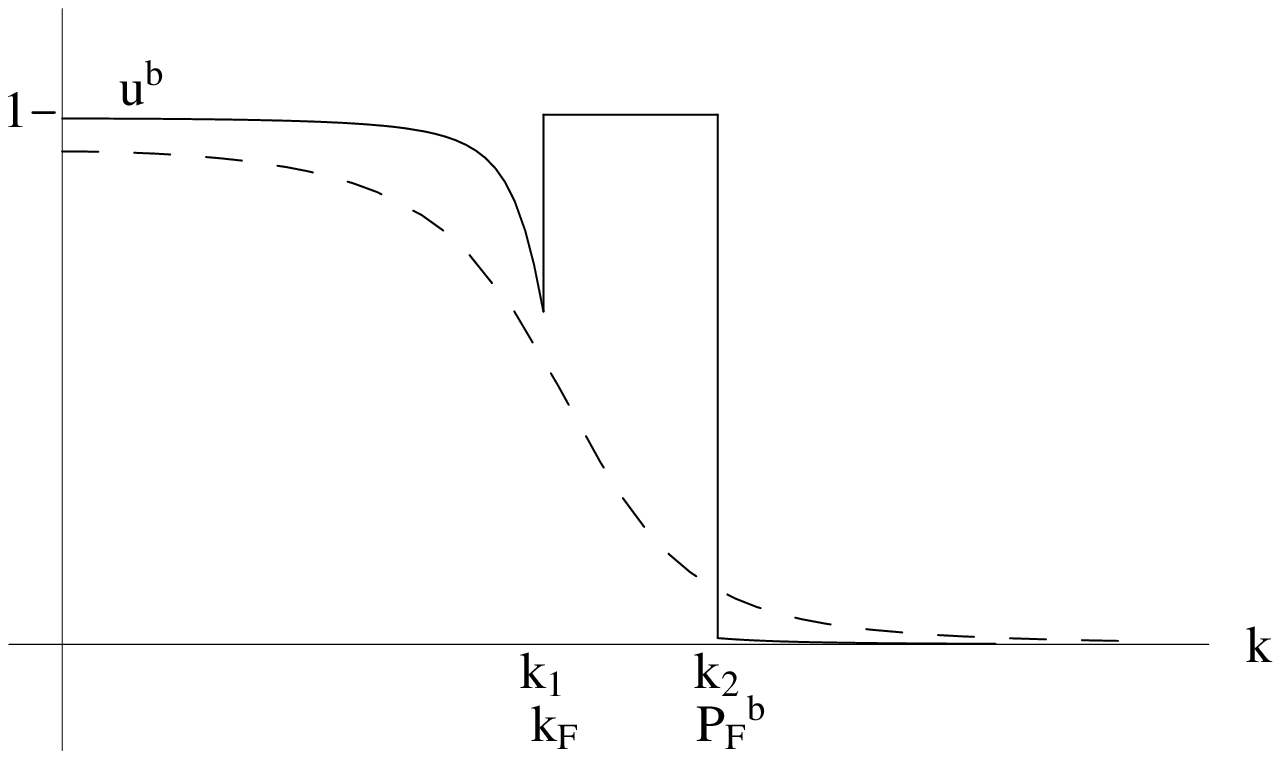}
\hspace{0.2in}
\includegraphics[height=1.5in]{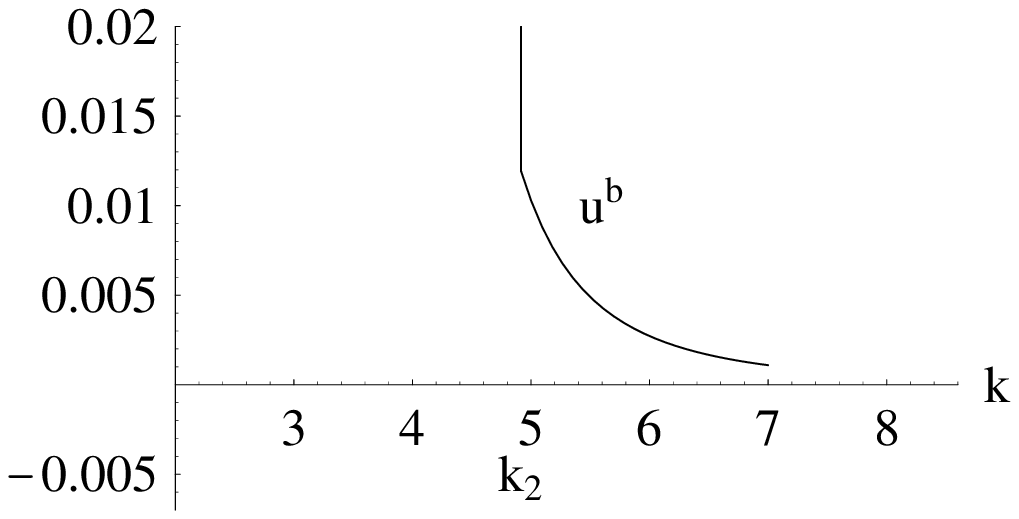}
\caption{\label{occupations}\textit{\protect Figures on top:
On the left it is shown the particle occupation number $u^a$ for the specie $a$ in the Sarma phase as compared with the BCS state (dashed curve). On right there is a amplified plot of the specie $a$ occupation number for $k \geq k_2$.
Figures on bottom: On left it is shown the particle occupation number $u^b$ for the specie $b$ in the Sarma Phase also compared with the BCS state (dashed curve). Again we show on right a amplified plot of the specie $b$ occupation number for $k \geq k_2$. We also show the hierarchy between the momenta involved, ${P_F^a}<k_1<k_F<k_2<{P_F^b}$.}}
\end{figure}

We show next that the solution for the gap equation in the Sarma phase can exist only for $g_{min}<g<g_{max}$. Since the integral in Eq.~(\ref{eq22}) is dominated for momenta around $k_F$, let us restrict
ourselves to the region $(k_F-\lambda) \leq |k| \leq k_1$ and $k_2 \leq |k| \leq (k_F+\lambda)$, where $\lambda$ is an ultraviolet cutoff
(used only for the sake of this proof, none of our results depend on it) and
suppose, for this analysis, that $P_F^b > P_F^a$ \cite{Liu1,Elena}.
$g_{min}$ is found when $\Delta$ has its maximum value namely, $\Delta_{max}=\Delta_{c}$.
\begin{equation}
\label{eq22-1}
|g_{min}|=\left[N(0) \ln \left(\frac{2~ \lambda~ k_F}{\delta p_F} \frac{(m_a + m_b)}{\sqrt{m_a m_b}} +
\sqrt{\left(\frac{2 ~\lambda~k_F}{\delta p_F} \frac{(m_a + m_b)}{\sqrt{m_a m_b}} \right)^2 + 1} \right) \right]^{-1},
\end{equation}
where $N(0)=\int \frac{d^3k}{(2 \pi)^3} \delta(\varepsilon_k^+) = \frac{M k_F}{ \pi^2}$ is the density of states.
The maximum value of $g$ is obtained when $\Delta=0$,
\begin{equation}
\label{eq22-2}
|g_{max}|=\left[N(0) \ln \left(\frac{2~ \lambda~ k_F}{\delta p_F} \frac{(m_a + m_b)}{\sqrt{m_a m_b}} \right) \right]^{-1}.
\end{equation}
A consequence of this is that for $\frac{2 ~\lambda~k_F}{\delta p_F} \frac{(m_a + m_b)}{\sqrt{m_a m_b}}>>1$
(or $\frac{m_b}{m_a} \to \infty$) the window between $g_{min}$ and $g_{max}$ is constant and independent of the cutoff $\lambda$
\begin{equation}
\label{eq22-3}
|g_{max}|^{-1} - |g_{min}|^{-1}=-\ln 2 N(0).
\end{equation}
Let us denote the gap of the standard BSC, which is given by Eq. (\ref{eq8}), by $\Delta_0$. We can find the gap in the Sarma phase through the identity
\begin{equation}
\label{eq23}
\frac{M}{2 \pi |a|}= \int \frac{d^3 k}{(2 \pi)^3} \frac{1}{\sqrt{{\varepsilon_k^+}^2+\Delta_0^2}}=\int_{\substack{k<k_1 \\ k>k_2}} \frac{d^3 k}{(2 \pi)^3}
\frac{1}{\sqrt{{\varepsilon_k^+}^2+\Delta^2}} .
\end{equation}
For small values of the gaps ($\Delta_0, \Delta << \mu_a, \mu_b$) the integrals can be approximated and it is found that \cite{Elena}
\begin{equation}
\label{eq24}
\frac{\Delta^2}{\Delta_0^2}=\frac{\varepsilon_k^a(k_1)}{\varepsilon_k^a(k_2)}.
\end{equation}
This equation for $\Delta$ can be expressed as
\begin{equation}
\label{eq25}
\frac{\Delta^2}{\Delta_0^2}=\frac{\delta p_F -[\delta p_F^2-16 m_a m_b \Delta^2]^{1/2}}
{\delta p_F +[\delta p_F^2-16 m_a m_b \Delta^2]^{1/2}}.
\end{equation}
It is easy to see that when $\Delta=\Delta_c$ we have one solution, $\Delta_S=\Delta_c=\Delta_0$,
which is the BCS case. The other solution for $\Delta$ is \cite{Sarma,Elena}
\begin{equation}
\label{eq26}
\Delta_S \simeq \sqrt{\Delta_0 \left( \frac{|\delta p_F|}{2 \sqrt{m_a m_b}}-\Delta_0 \right)}.
\end{equation}
This equation for $\Delta$ generalizes the Sarma results \cite{Sarma}. The conclusion is that
\begin{eqnarray}
\label{eq27}
2 \sqrt{m_a m_b}\Delta_0& \leq |\delta p_F| \leq &4 \sqrt{m_a m_b}\Delta_0\\
\nonumber
0&\leq\Delta_S\leq&\Delta_0.
\end{eqnarray}
The number densities in the Sarma phase are given by $n_{\alpha} = -\frac{\partial {\rm F}^S}{\partial \mu_{\alpha}}$, then we get
\begin{align}
\label{eq28}
n_a &=\int_{\substack{k<k_1 \\ k>k_2}} \frac{d^3k}{(2 \pi)^3} \frac{1}{2} \left(1-\frac{{\varepsilon}_k^+}{\sqrt{{\varepsilon_k^+}^2+\Delta^2}} \right) ,\\
\nonumber
n_b &=n_a+\int_{k_1}^{k_2} \frac{d^3k}{(2 \pi)^3}.
\end{align}
It shows that outside the region $k_1<k<k_2$ the particles $a$ and $b$ are paired via BCS having number density $n_a$, whereas inside the region of momenta $k_1<k<k_2$ there are only $b$ particles. We define the particle occupation numbers for the species $a$ and $b$  respectively, as
\begin{align}
\label{occup}
u^a &=\frac{1}{2} \left(1-\frac{{\varepsilon}_k^+}{\sqrt{{\varepsilon_k^+}^2+\Delta^2}} \right),~~{\rm for}~~ k<k_1~~ {\rm and} ~~k>k_2, \\
\nonumber
u^b &=\frac{1}{2} \left(1-\frac{{\varepsilon}_k^+}{\sqrt{{\varepsilon_k^+}^2+\Delta^2}} \right),~~{\rm for}~~ k<k_1~~ {\rm and} ~~k>k_2 ,~~ {\rm and}~~ 1~~ {\rm for}~~ k_1<k<k_2.
\end{align}
In Fig. (\ref{occupations}) we show the particle occupation numbers for the species $a$ and $b$. We also plot the hierarchy of momenta for those values of the parameters we have used in Fig. (\ref{fig1}). As we can see, the number of particles of both species having momenta $k>k_2$ is negligible. This permits us to say that the approximate number density of particles available to form pairs in the Sarma phase is
\begin{equation}
\label{occup2}
n^S \approx \int_0^{k_1} \frac{dk ~k^2}{2 \pi^2} \frac{1}{2} \left(1-\frac{{\varepsilon}_k^+}{\sqrt{{\varepsilon_k^+}^2+\Delta^2}} \right)~<~n^{BCS}=
\int_0^{\infty} \frac{dk ~k^2}{2 \pi^2} \frac{1}{2} \left(1-\frac{{\varepsilon}_k^+}{\sqrt{{\varepsilon_k^+}^2+\Delta_0^2}} \right).
\end{equation}
This fact can have consequences in the condensation energy of the Sarma phase since, as we have seen in Eq. (\ref {cond}), pairing lowers the energy of the system.

\section{Searching for the Lowest Energy}
\subsection{Fixed chemical potentials}
The situation where the chemical potentials are kept fixed can find place for instance in a gas of fermionic atoms connected to reservoirs of species $a$ and $b$ so the number densities are allowed to change in the system.

\begin{figure}[t]
\includegraphics[height=2.2in]{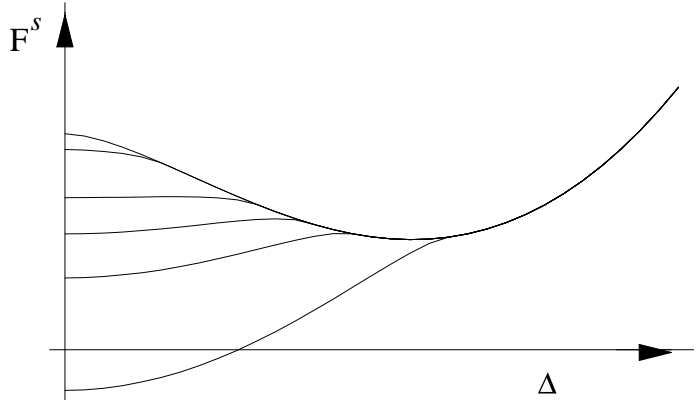}
\caption{\label{omega}\textit{thermodynamic potential ${\rm F}^{S}$ for different values of
$P_F^b$ and $P_F^a$ ($k_F$ fixed). The top curve corresponds to $P_F^a=P_F^b$ and
the lower curves correspond to increasing values of $|P_F^b-P_F^a|$.} }
\end{figure}

In Fig.~\ref{omega} we show the thermodynamic potential ${\rm F}^{S}$ as a function of
$\Delta$ for different values of $P_F^a$ and $P_F^b$,  keeping the combination
$k_F^2/2M={P_F^a}^2/2m_a + {P_F^b}^2/2m_b$ fixed, computed from a numerical evaluation
of Eq. (\ref{eq21}).
When $\Delta_S$ is larger than $\Delta_c=\frac{|\delta p_F|}{4\sqrt{m_a m_b}}$, $k_{1,2}^2$ are not real, the quasi-particle excitations ${\cal E}_k^{\alpha,\beta}$ are always positive, and the free-energy is unchanged from the $P_F^a=P_F^b$ situation. However, for $\Delta_S < \Delta_c$, $k_{1,2}^2$ are positive and the free-energy can be lowered by filling the states between $k_1$ and $k_2$ with $b$ particles. The $\Delta=\Delta_0$ corresponds to the BCS phase. The normal state ($\Delta=0$), in particular, can have its free-energy lowered with increasing $P_F^b-P_F^a$, and after $P_{F, DM}^b-P_F^a$ (see discussion on $P_{F, DM}$ below) becomes smaller than the previous minimum at $\Delta_0$. Then, for fixed $k_F$ and increasing $P_F^b-P_F^a$, there is a first-order phase transition between the superfluid and the normal state, as it has been found in various physical situations \cite{Norman1,Sedrakian,Paulo}.

We can understand this results with very simple physical terms in the following way. Starting with $\mu_a=\mu_b$, we have the BCS state with its free-energy given by Eq. (\ref{eq11})
\begin{equation}
\label{pot1}
{\rm F}(\mu_a= \mu_b=\mu)={\rm E}(n) - (\mu_a  +\mu_b) n ={\rm E}(n) - \mu_{Total}~n,
\end{equation}
where $\mu_{Total}=2\mu=\frac{k_F^2}{2M}$ is kept fixed.

In order to compare the energies of the BCS state with another state having $\mu_b > \mu_a$ ($P_F^b > P_F^a$), we let the system absorb one $b$ particle from the $b$-particles reservoir\footnote{This will increase the energy of the system by at least the value of $\Delta_0$. The extra $b$ particle added will be located at the level of elementary excitations.} and eliminate one $a$ particle to the $a$-particles reservoir (Again this will increase the energy of the system by at least $\Delta_0$, by breaking a pair and promoting the remaining $b$ particle to the elementary excitation levels). Then, the free-energy of the system having unequal Fermi surfaces turns out to be
\begin{align}
\label{pot2}
{\rm F}(\mu_b > \mu_a)&={\rm E}(n)+2 \Delta_0 - \mu_a (n-1) -\mu_b (n+1) \\
\nonumber
&= {\rm F}(\mu_a= \mu_b) + 2 \Delta_0 -(\mu_b- \mu_a),
\end{align}
where we have used that $\mu_a+\mu_b=\mu_{Total}$. By absorbing one $b$ particle and eliminating one $a$ particle the system reduces its free-energy due to the $(\mu_b- \mu_a)$ term but also increases ${\rm F}$ due to the $2 \Delta_0$ term. Then, this is energetically favorable only if the difference in chemical potentials is large enough or the gap is small enough. In spit of the mismatch in the chemical potentials, the BCS state with equal number of $a$ and $b$ particles remains the ground state while
\begin{equation}
\label{pot3}
\Delta_0 > \delta \mu,
\end{equation}
where $\delta \mu = \frac{ \mu_b- \mu_a}{2}$.

As one can sees in Fig.~\ref{omega}, there exist a special combination of $P_F^a$ and $P_F^b$ for which ${\rm F}^S$ has double minima. Then, given a $P_F^a$, we want to know the correspondent value of $P_F^b$ that satisfies this requirement. We note that when $\Delta=0~\to~k_1=P_F^a$ and $k_2=P_F^b$ whereas for $\Delta=\Delta_0~\to~k_1=k_2=\sqrt{\frac{{P_F^a}^2+{P_F^b}^2}{2}}$. Thus, the condition to find the relation between $P_F^a$ and $P_F^b$ is
\begin{align}
\label{min1}
& {\rm F}(\Delta=0)={\rm F}(\Delta=\Delta_0) \to \\
\nonumber
& \int_{\substack{k<P_F^a \\ k>P_F^b}} \frac{d^3k}{(2 \pi)^3} ({\varepsilon}_k^+ - |{\varepsilon}_k^+|)+
\int_{P_F^a}^{P_F^b} \frac{d^3k}{(2 \pi)^3} {\varepsilon}_k^b  =
\int \frac{d^3k}{(2 \pi)^3} ({\varepsilon}_k^+ - E)-\frac{\Delta_0^2}{g}.\
\end{align}
Since ${\rm F}$ is being evaluated at its minimum, we can substitute the r.h.s. of the equation above by Eq. (\ref{eq3}) (whose solution is given by Eq. (\ref{eq9})) and the l.h.s. is just the free-energy of the unpaired $a$ and $b$ particles
\begin{eqnarray}
\label{min2}
-\frac{1}{30 \pi^2}\left[\frac{{P_F^a}^5}{m_a}+\frac{{P_F^b}^5}{m_b}\right]=-\frac{k_F^5}{30 \pi^2 M}-\frac{M k_F }{2 \pi^2} \Delta_0^2.
\end{eqnarray}
The solution of the equation above we cal $P_{F, DM}^b$, where $DM$ stands for double minima. We point out here that this situation means coexistence of phases. Given these fixed values of $P_F^b$ and $P_F^a$ (or $\mu_a$ and $\mu_b$) found above, the system could either be found in the normal, BCS or in a mixed phase, since those states would have the same minimal energy. When the free-energy ${\rm F}(P_{F, DM}^b)$ has its minimum at $\Delta_0$,  after a increase in $P_F^b$ the gap jumps from $\Delta_0$ to $0$, characterizing a first order phase transition from the superfluid to the normal phase, as we have discussed already.

According to Eq.~(\ref{eq27}) there is (for some values of the Fermi surfaces $P_F^a$ and $P_F^b$) an unstable state, that is the Sarma phase we have discussed in the previous section, in addition to the stable (or metastable) normal and BCS phases. This state corresponds to a maximum of $\rm F$ as a function of $\Delta$ situated between the BCS minimum $\Delta_0$ and the normal phase at $\Delta=0$.

\begin{figure}[t]
\includegraphics[height=1.8in]{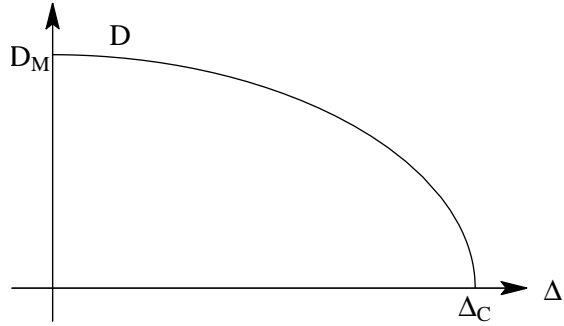}
\caption{\label{dist}\textit{Range of momenta allowed for the excess of particles of specie $b$ in the Sarma phase, where $0 \leq \Delta \leq \Delta_c$.}}
\end{figure}
We can cast some light in the understanding of Figures~\ref{occupations} and~\ref{omega} (and consequently learn what really happens in the situation of fixed chemical potentials in the Sarma phase ($0 \leq \Delta \leq \Delta_c$)). With that purpose we define a quantity ${\rm D}(\Delta)=k_2-k_1$ which is the region of momenta allowed for the (unpaired) $b$ particles in the Sarma phase. For any curve having momenta $P_F^b>P_F^a$ (and fixed in each curve) from the second curve of Fig.~\ref{omega} until $P_{F, DM}^b$ the minimum of the free-energy of the system is achieved in the following way. When the (weak) attraction is turned on and increases, the system keeps releasing the ``excess" of $b$ particles until the point ${\rm D}(\Delta_c)=0$ (see Fig.~\ref{dist}) where the BCS condition is reached and both species have the same density $n_a=n_b=\frac{k_F^3}{6 \pi^2}+\frac{M^2 \Delta_c^2}{2 \pi^2 k_F} \left(5 + \frac{\pi}{ |a| k_F} \right)$. For $P_F^b>P_{F, DM}^b$ the minimum of {\rm F} is obtained when the attractive interaction between the fermions decreases until it is turned off (${\rm D}(\Delta=0) \equiv {\rm D}_{\rm M}={P_F^b}-{P_F^a}$). There is absorption of particles from the reservoirs so the asymmetry in the number densities gets its maximum and the system is in the normal state with number densities $n_a=\frac{{P_F^a}^3}{6 \pi^2}$ and $n_b=\frac{{P_F^b}^3}{6 \pi^2}$.

\subsection{Fixed particle number}
The analysis regarding the ground state of asymmetrical fermion superfluids we had so far is valid for systems where the chemical potentials of the different species are kept fixed. We turn now to the cases in which the particle numbers are the fixed quantities. This is the relevant situation for trapped cold fermionic atoms. As we have seen, in the BCS state the particles have the same density, whereas in the normal ground state the number densities can be different. A mixed (inhomogeneous) phase has been proposed in Ref. \cite{PRL1} as being composed by islands of normal matter in a sea of BCS phase. According to Eq. (\ref{eq28}) the Sarma phase also supports different particle numbers, depending on the value of the gap. In order to find the ground state of a system with fixed $n_a$ and $n_b$ number densities it is, then, natural to compare the energies of the normal, Sarma and mixed phases.

\subsubsection{The mixed phase energy}
In the mixed phase, $n_a$ and $n_b$ particles in a box with volume $V$, are accommodated in such a way that in a fraction $x$ of this volume the particles are free having densities $\tilde n_{a,b}$, and in the rest of the volume there is pairing formation between the species $a$ and $b$ with number densities $n_a^{\rm BCS}=n_b^{\rm BCS}=n$. Then, the number densities in each component of the mixed phase read
\begin{eqnarray}
\label{eq0}
n_a =   x\tilde n_a + (1-x)n,\\
\nonumber
n_b = x\tilde n_b + (1-x)n,
\end{eqnarray}
with $0 \leq x \leq 1$. The preferable mixed state for given $n_a$ and $n_b$ particle densities is the one which has the lowest energy
\begin{equation}
\label{eq13}
{\rm E}^{\rm MIX}(n_a,n_b)=Min_{x, n}\left\{ x {\rm E}^{\rm N}(\tilde n_a,\tilde n_b)+ (1-x){\rm E}^{\rm BCS}( n) \right\},
\end{equation}
where ${\rm E}^{\rm N}$, the energy of the normal (unpaired) particles, is given by

\begin{align}
\label{eq1}
{\rm E}^{\rm N}(\tilde n_a,\tilde n_b)&=<{\rm BCS}|H(g=0)|{\rm BCS}>=\sum_{k < P_F^{\alpha},\alpha}
\xi_k^{\alpha}=\frac{(6\pi^2)^{\frac{5}{3}}}{20\pi^2}
\left[ \frac{\tilde n_a^{5/3}}{m_a}+ \frac{\tilde n_b^{5/3}}{m_b} \right]\\
\nonumber
&=\frac{(6\pi^2)^{\frac{5}{3}}}{20\pi^2}
\left[ \left(\frac{n_a-(1-x)n}{x}\right)^{5/3} \frac{1}{m_a}+\left(\frac{n_b-(1-x)n}{x}\right)^{5/3} \frac{1}{m_b} \right],
\end{align}
and ${\rm E}^{\rm BCS}(n)$ is given by Eq. (\ref{eq12}).
We neglect the interface energy between the normal and superconductor components of the mixed phase Fermi gas.
For large enough systems the surface tension effects are small. Consequently, the mixed phase is energetically favorable (and stable)
if, and only if, the gain in bulk free energy is larger than the disregarded surface energy \cite{Buballa}.

\subsubsection{On the conditions for the existence of equilibrium between the phases}

The criteria for thermodynamical equilibrium between the components of a mixed phase are dictated by the Gibbs conditions
\begin{eqnarray}
\label{crit}
{\rm I.}~~{\rm T^N=T^S},\\
\nonumber
{\rm II.}~~\mu^{\rm N}=\mu^{\rm S},\\
\nonumber
{\rm III.}~~{\rm P^N = P^{S}}.\\
\nonumber
\end{eqnarray}
These conditions ensure thermal, chemical and mechanical equilibrium, respectively between the normal and superconducting phases.
Those requirements are enforced  by the Maxwell construction we employ. Condition I is automatically satisfied since we are considering a cold system.
As we show below, conditions II and III are reached when we find $n$ and $x$ that minimizes the energy. At the minimum we write
\begin{align}
\label{cond1}
\frac{\partial }{\partial n}{\rm E}^{\rm MIX}(n_a,n_b)&=x \frac{\partial}{\partial n}{\rm E}^{\rm N}(\tilde n_a,\tilde n_b) +
 (1-x) \frac{\partial}{\partial n}{\rm E}^{\rm BCS}(n) \\
\nonumber
& = x \frac{\partial}{\partial n}[{\rm F}^{\rm N}(\tilde \mu_a^{\rm N},\tilde \mu_b^{\rm N})+ \tilde \mu_a^{\rm N}~\tilde n_a +
\tilde \mu_b^{\rm N}~\tilde n_b] + (1-x) \frac{\partial}{\partial n}[{\rm F}^{\rm BCS}(\mu)+ \mu n]=0 \to \\
\nonumber
& \tilde \mu_a^{\rm N} + \tilde \mu_b^{\rm N}=\mu,
\end{align}
where $\mu=\mu_a^{\rm BCS} + \mu_b^{\rm BCS}$, $\tilde \mu_{\alpha}^{\rm N}=\frac{(6 \pi^2 \tilde n_{\alpha})^{2/3}}{2 m_{\alpha}}$
and $\tilde n_{\alpha} =
\frac{n_{\alpha}-(1-x)n}{x}$. This result tells us that the sum of the chemical potentials between the normal and superconducting phases in
equilibrium is the same, and not the individual ones. This is because the BCS phase can exchange with the normal phase only pairs of $a$ and $b$
species to reach the equilibrium. For the mechanical equilibrium, we have
\begin{align}
\label{cond2}
\frac{\partial }{\partial x}{\rm E}^{\rm MIX}(n_a,n_b) &=  {\rm E}^{\rm N}(\tilde n_a,\tilde n_b)-{\rm E}^{\rm BCS}(n) +
x \frac{\partial}{\partial x}{\rm E}^{\rm N}(\tilde n_a,\tilde n_b)=0 \to \\
\nonumber
{\rm F}^{\rm N}(\tilde \mu_a^{\rm N},\tilde \mu_b^{\rm N}) &={\rm F}^{\rm BCS}(\mu),
\end{align}
where we have used the result of Eq. (\ref{cond1}). This is not surprising. We have found already in the situation of fixed chemical potentials that
coexisting phases have the same thermodynamic potential. Since the thermodynamic potential is related to the pressure by ${\rm F}=-{\rm P}$, we obtain
the desired result ${\rm P^N(\tilde \mu_a^{\rm N},\tilde \mu_b^{\rm N})=P^{BCS}(\mu_a^{\rm BCS} + \mu_b^{\rm BCS})}$. Then we have found that the mixed phase state
satisfies all the necessary conditions to be in thermodynamical equilibrium.

\subsubsection{The Sarma phase energy}

With Eqs. (\ref{eq21}) and (\ref{eq28}), we can write the energy in the Sarma phase as
\begin{align}
\label{es}
{\rm E}^{\rm S} &={\rm F}^{\rm S}+\sum_{\alpha} \mu_{\alpha} n_{\alpha}\\
\nonumber
&=\int_{\substack{k<k_1 \\ k>k_2}} \frac{d^3k}{(2 \pi)^3} [{\varepsilon}_k^+ -E + \mu~v_k^2 ]+
\int_{k_1}^{k_2} \frac{d^3k}{(2 \pi)^3} \frac{k^2}{2 m_b} -\frac{\Delta^2}{g}.
\end{align}
Just for completeness, we show in Section \ref{ape} the expressions for the energy and free energy of the Sarma phase and its (expected) results for the limits $\Delta \to 0$ and $\Delta \to \Delta_0$.

\subsubsection{The lowest energy}

Using the conditions for equilibrium between the normal and superfluid phases, Eqs. (\ref{cond1}) and (\ref{cond2}), the mixed phase energy can be rewritten at the minimum as

\begin{align}
\label{mixen2}
{\rm E}_{Min}^{\rm MIX}(n_a,n_b) &= x [{\rm F}^{\rm N}(\tilde \mu_a^{\rm N},\tilde \mu_b^{\rm N})+\tilde \mu_a^{\rm N}~\tilde n_a + \tilde \mu_b^{\rm N}~\tilde n_b] + (1-x)[{\rm F}^{\rm BCS}(\mu)+ \mu n] \\
\nonumber
&={\rm F}^{\rm BCS}(\mu) + \tilde \mu_a^{\rm N}~n_a +\tilde \mu_b^{\rm N}~n_b,
\end{align}
where the $n$ and $x$ dependences in $\tilde \mu_a^{\rm N}$ and $\tilde \mu_b^{\rm N}$, are the solutions of Eqs. (\ref{cond1}) and (\ref{cond2}), and should be termed $n_{Min}$ and $x_{Min}$. Both parameters shall depend only, of course, $m_a,~ m_b,~ n_a,~ n_b$ and $a$.

Next we define a function ${\Delta {\rm E}(n_a,n_b)}$ as the energy difference between the mixed and the Sarma phases
\begin{align}
\label{eq31}
{\Delta {\rm E}}(n_a,n_b)&={\rm E}_{Min}^{\rm MIX}(n_a,n_b)-{\rm E}^{\rm S}(n_a,n_b) \\
\nonumber
&= {\rm F}^{\rm BCS}(\mu) + \tilde \mu_a^{\rm N}~n_a + \tilde \mu_b^{\rm N}~n_b - [{\rm F}^{\rm S}(\mu_a^{\rm S},\mu_b^{\rm S}) + \mu_a^{\rm S}~n_a + \mu_b^{\rm S}~n_b]\\
\nonumber
&={\rm F}^{\rm BCS}(\mu)-{\rm F}^{\rm S}(\mu_a^{\rm S},\mu_b^{\rm S}) + n_a (\tilde \mu_a^{\rm N}-\mu_a^{\rm S}) +
n_b (\tilde \mu_b^{\rm N}-\mu_b^{\rm S}).
\end{align}

Now we are able to compare the energy of these two phases for those values of the gap supported by the Sarma phase. As we have seen in the situation of
fixed chemical potentials, the energy difference ${\rm F}^{\rm S}(\Delta_0)-{\rm F}^{\rm S}(\Delta_S)$ is always negative for $P_{F}^b < P_{F, DM}^b$
(for fixed $P_{F}^a$). For $P_{F}^b>P_{F, DM}^b$, the mismatch in the Fermi surfaces is large enough so the ground state is the normal phase.
For the case of $a$ and $b$ fermionic atoms confined in an atomic trap we investigate below its ground state in three possibilities that could be
implemented in current experiments. In the first two cases we vary the interaction through the scattering length $a$ and in the last one we fix $a$
and vary the number densities. In most of the situations ${\Delta {\rm E}}(n_a,n_b)$ has to be evaluated numerically but, as we are going to see next, in some limiting cases we can obtain analytical results.
\newline

{\it 4.1:}~{\it Fixed $n_a$ and $n_b$ with ${\Delta {\rm E}}(n_a,n_b,\Delta_S=0)$}. When $\Delta_S \to 0$, which corresponds to the lower limit for the Fermi surfaces mismatch in Eq.~(\ref{eq27}), ${P_F^b}^2-{P_F^a}^2=2 \sqrt{m_a m_b}\Delta_0$, we have ${\rm F}^{\rm S} \to {\rm F}^{\rm N}$
(see Eq. (\ref{eq32-2}) in the appendix) and consequently $\mu_{\alpha}^{\rm S} \to \mu_{\alpha}^{\rm N}$. Then
\begin{equation}
\label{res1}
{\Delta {\rm E}}(n_a, n_b,\Delta_S=0) = {\rm F}^{\rm BCS}(\mu)-{\rm F}^{\rm N}(\mu_a^{\rm N},\mu_b^{\rm N}) + n_a (\tilde \mu_a^{\rm N}-\mu_a^{\rm N}) +
n_b (\tilde \mu_b^{\rm N}-\mu_b^{\rm N}),
\end{equation}
where $\mu_{\alpha}^{\rm N}=\frac{(6 \pi^2 n_{\alpha})^{2/3}}{2 m_{\alpha}}$.

For very large asymmetry between the number densities $n_a$ and $n_b$, $x_{Min}$ approaches 1 which implies $\tilde \mu_{\alpha}^{\rm N} \to \mu_{\alpha}^{\rm N}$ so the mixed phase would also be reduced to normal phase resulting in ${\Delta {\rm E}}(n_a,n_b \gg n_a,\Delta_S=0)=0$. This happens because with large asymmetry the Fermi surfaces are separated apart, making pairing energetically disfavored and the preferred ground state is the normal phase.

We prove analytically below that ${\Delta {\rm E}}(n_a,n_b,\Delta_S=0)$ is negative for small
asymmetry on the number densities of the $a$ and $b$ particle species. In this case the Sarma energy can be written as
\begin{align}
 {\rm E}^{\rm S}(n_a,n_b,\Delta_S=0) &=\frac{(6\pi^2)^{5/3}}{20\pi^2}
\left[ \frac{n_a^{5/3}}{m_a}+\frac{(n_a+\delta n)^{5/3}}{m_b}   \right]\nn\\
&\cong\frac{(6\pi^2n_a)^{5/3}}{20\pi^2 M}\[1
+\frac{5 M}{3 m_b}\frac{\delta n}{n_a}
     +\frac{5 M}{9m_b}\left(\frac{\delta n}{n_a}\right)^2+\mathcal{O}(\delta n^3/n_a^3)\],
\end{align}
where $\delta n = n_b-n_a$ is assumed to be small, $\delta n\ll n_a$.

An upper bound on ${\Delta {\rm E}}(n_a,n_b,\Delta_S=0)$ can be obtained by setting the density of the
superconducting component of the mixed phase $n=n_a$ and minimizing with respect to $x$. Then we have:
 \begin{align}
 \label{mix-sarma}
 {\Delta {\rm E}}(n_a,n_b,\Delta_S=0) &\cong -(1-x)(6\pi^2 n_a)^{1/3} \frac{M \Delta_0^2(P_F^a)}{2\pi^2}\\
  \nonumber
 &+ \frac{(6\pi^2 n_a)^{5/3}}{36\pi^2 m_b}\frac{\delta n^2}{n_a^2}\left( \frac{1}{x}-1\right)
 +\mathcal{O}(n_a^{-4/3}\delta n^3),
 \end{align}  whose minimization yields
  \begin{align}
  \label{mix-sarma2}
 x = x_{Min} &= \sqrt{\frac{(6\pi^2 n_a)^{4/3} }{18 M m_b \Delta_0^2(p_F^a)}}  \frac{\delta n}{n_a}.
 \end{align}
Thus, the energy difference read
\begin{align}
  \label{mix-sarma3}
 {\Delta {\rm E}}(n_a,n_b,\Delta_S=0) &\cong -  (6\pi^2 n_a)^{1/3} \frac{M \Delta_0^2(p_F^a)}{2\pi^2} (1-x_{Min})^2 < 0.
 \end{align}
The limit $\delta n \to 0$ implies $x_{Min} \to 0$, and ${\rm E}^{\rm MIX}$ is entirely in the BCS phase. Numerical calculations show that the upper bound above is close to the actual minimum. Eq.~(\ref{mix-sarma2}) shows that the mixed phase is energetically favored compared to the Sarma phase in one extreme of the window in Eq.~(\ref{eq27}).
\newline

{\it 4.2:}~{\it Setting the Sarma gap in its maximum value: ${\Delta {\rm E}}(n_a,n_b, \Delta_S = \Delta_0)$}. This situation corresponds to the upper limit for the Fermi surfaces mismatch in Eq.~(\ref{eq27}), ${P_F^b}^2-{P_F^a}^2=4 \sqrt{m_a m_b}\Delta_0$. In this case, $\Delta_S \to \Delta_0$, ${\rm F}^{\rm S} \to {\rm F}^{\rm BCS}$ (see appendix), $n_a=n_b=n$, $(\mu_a^{\rm S}+\mu_b^{\rm S}) \to (\mu_a^{\rm BCS}+\mu_b^{\rm BCS})$ and the energy difference in Eq.~(\ref{eq31}) is
\begin{equation}
\label{ res2}
{\Delta {\rm E}}(n_a=n_b=n) = n [\tilde \mu_a^{\rm N}+\tilde \mu_b^{\rm N}-(\mu_a^{\rm BCS}+\mu_b^{\rm BCS})]=0,
\end{equation}
since for ${\rm E}^{\rm MIX}(n_a,n_b)$, from Eq. (\ref{cond1}), we have $\tilde \mu_a^{\rm N}+\tilde \mu_b^{\rm N}=\mu_a^{\rm BCS}+\mu_b^{\rm BCS}$.
This means that in this situation both the mixed and the Sarma phases are in the BCS state, having the same energy.
\newline

\begin{figure}[t]
\includegraphics[height=1.9in]{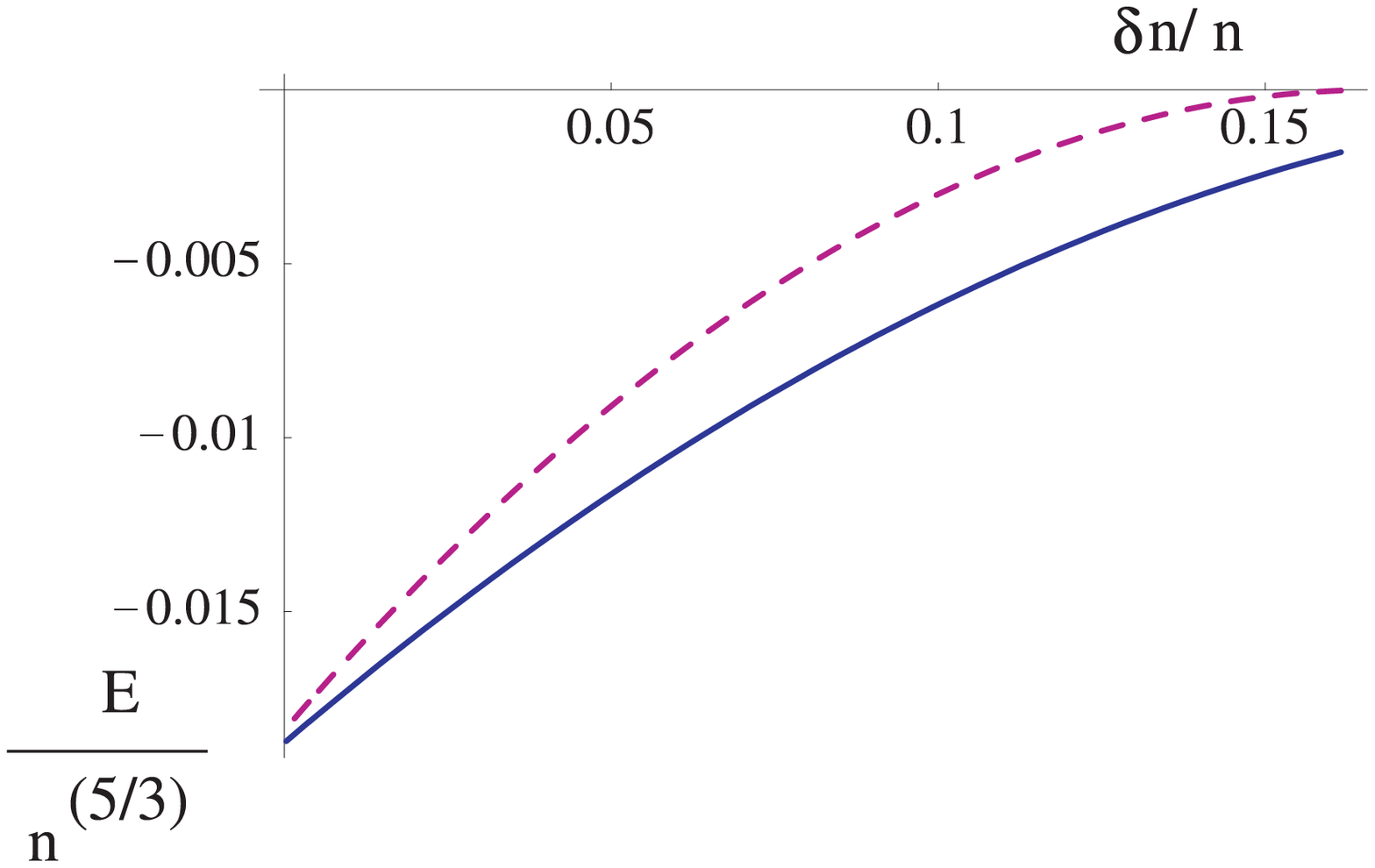}
\hspace{0.2in}
\includegraphics[height=1.7in]{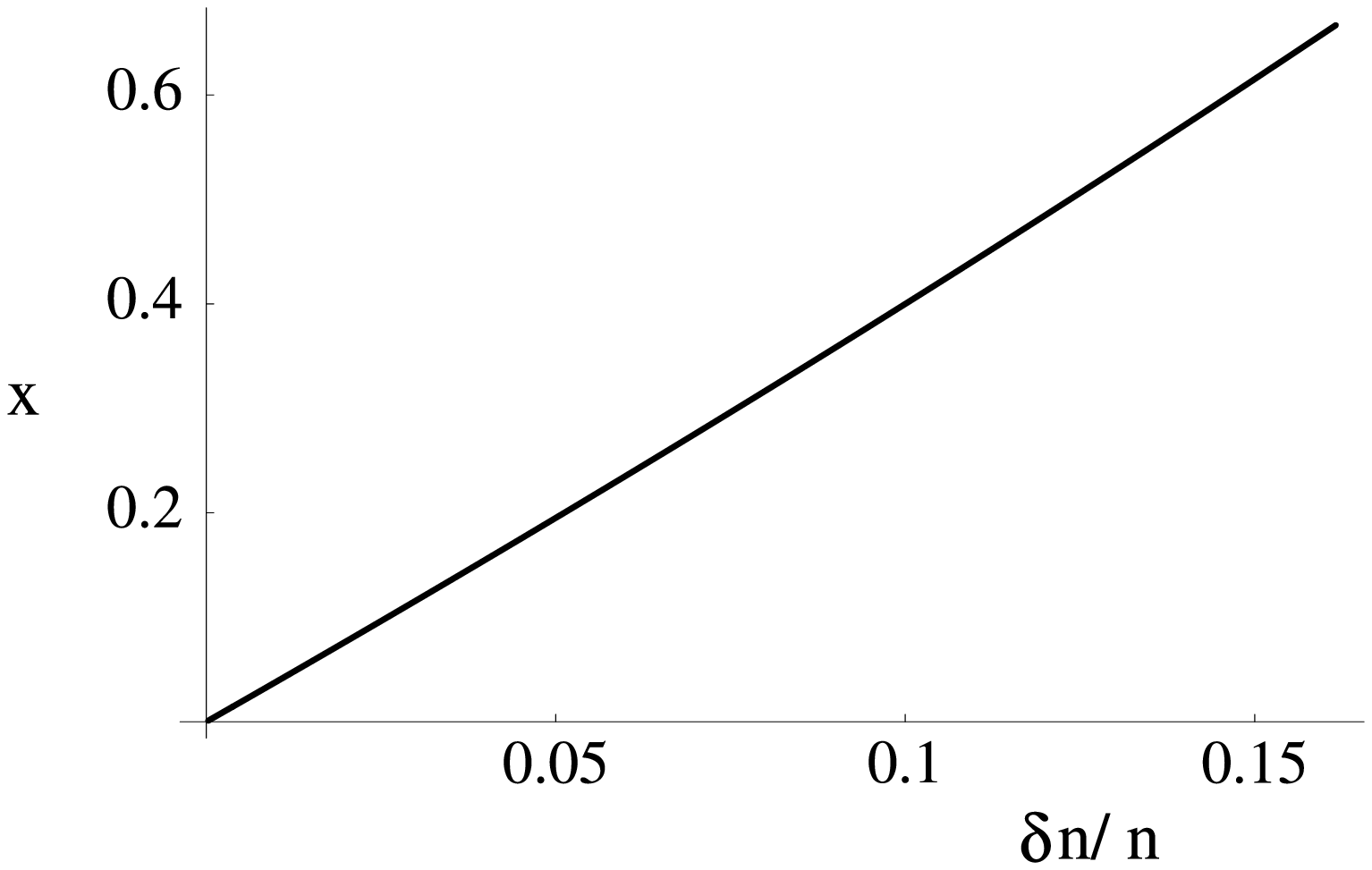}
\caption{\label{mix_and_sarma}\textit{\protect Figure on the left:
Energy of the Sarma (dashed curve) and
mixed phase (solid curve) relative to the normal phase as a function of $\delta
n/n$ where $n=(n_A+n_B)/2$ with $M_B=7M_A$.
The values of $n_A$ and $n_B$ were obtained from the Sarma phase for a fixed $a$,
and  $|a|p_A=0.59$, $0.63\le |a| p_B\le 0.65$.
Figure on the right: The fraction of normal state
$x$ as a function of $\delta n/n$ with the same masses and scattering length.}
 }
\end{figure}

{\it 4.3:}~{\it Fixed scattering length varying the number densities with ${\Delta {\rm E}}(\delta n)$}. For intermediate values of $P_F^b-P_F^a$ and
for fixed scattering length $a$ (still satisfying the constraint in Eq. (\ref{eq27})), the difference $\Delta{\rm E}(n_a, n_b)$ interpolates between
the extremes in {\it 4.1} and {\it 4.2}, as Fig.~(\ref{mix_and_sarma}) exemplifies. We find that for all reasonable  values of the parameters
(that is, where the mean field analysis should apply), and for fixed particle numbers $n_a$ and $n_b$, the mixed phase has a smaller energy
than the Sarma phase.

\subsection{Fixed overall density}

We discuss now another case that happens when the overall number of particles $N=n_a + n_b$ is fixed but conversions between the $a$ and $b$ species,
$a \leftrightarrow b$, are allowed to occur. This situation is relevant, for instance, in astro-particle physics. The core of neutron stars is believed to be formed by
high density quark matter, where different quarks interchange their flavors because of weak interactions. In this case, the thermodynamic function to
be minimized is \cite{PRL1}

\begin{align}
\label{k1}
K(N=n_a+n_b)&={\rm E}^{\rm MIX}(N) -x \delta \tilde \mu^{\rm N}~\delta \tilde n^{\rm N} - (1-x) \delta \mu^{\rm BCS}~\delta n^{\rm BCS}  \\
\nonumber
&= {\rm E}^{\rm MIX}-\delta \tilde \mu^{\rm N} (n_b - n_a) ,
\end{align}
where $\delta \tilde \mu^{\rm N} = \frac{\tilde \mu_b^{\rm N}-\tilde \mu_a^{\rm N}}{2}$, $\delta \tilde n^{\rm N} = \tilde n_b - \tilde n_a$,
$\delta \mu^{\rm BCS} = \frac{\mu_b^{\rm BCS}-\mu_a^{\rm BCS}}{2}$, $\delta n^{\rm BCS}=0$ and  we have made use of Eq. (\ref{eq0}).

The particles in the normal component of the mixed phase are accommodated in such a way that its energy is given by
\begin{equation}
\label{k2}
{\rm E}^{\rm N}(\tilde N)=\frac{(6 \pi^2)^{5/3}}{20 \pi^2}
\left[ \frac{( \frac{\tilde N}{2}-z)^{5/3}}{m_a} + \frac{( \frac{\tilde N}{2}+z)^{5/3}}{m_b}\right],
\end{equation}
where $\tilde N=\tilde n_a + \tilde n_b=\frac{N-2(1-x)n}{x}$ is the fixed overall number density in the normal fraction of the mixed phase.
Minimization with respect to $z$ yields
\begin{equation}
\label{k3}
{\rm E}^{\rm N}(\tilde N)=\frac{(6 \pi^2 \tilde N)^{5/3}}{20 \pi^2} \frac{1}{(m_a^{3/2}+m_b^{3/2})^{2/3}}.
\end{equation}
Then, the most favored mixed state for fixed total density $N$ is the one with the smallest ``$K$ energy'':
\begin{align}
\label{k4}
K^{\rm MIX}(N=n_a+n_b)&=\text{Min}_{x, n}\left\{ x \frac{(6 \pi^2 )^{5/3}}{20 \pi^2} \frac{1}{(m_a^{3/2}+m_b^{3/2})^{2/3}}
\left( \frac{N-2(1-x)n}{x} \right)^{5/3} \right.\nn\\
&{}\left.+(1-x) \left[ \frac{(6\pi^2  n)^{5/3}}{20\pi^2 M}-
\frac{M\Delta_0^2(n) (6\pi^2 n)^{1/3}}{2\pi^2} \right] -\delta \tilde \mu^{\rm N} (n_b - n_a) \right\}.
\end{align}
The two interesting limits that should be evaluated in the situation of fixed overall density are $m_a=m_b$ and $m_b \gg m_a$.\\
\newline
{\bf I:}~$m_a=m_b$. We set $\delta \tilde \mu^{\rm N}=0$ to ensure the equilibrium in the conversions between the $a$ and $b$ species
in the (normal) subsystem of the mixed phase\footnote{The same equilibrium is established, of course, in the superfluid component of the mixed phase, i.e. $\delta \mu^{\rm BCS}=0$.}.
The chemical equilibrium implies $\frac{{\tilde n_a}^{2/3}}{m_a}=\frac{{\tilde n_a}^{2/3}}{m_b}$. Thus, we obtain
\begin{equation}
\label{k5}
n_a=(1-x)n+(\frac{m_a}{m_b})^{3/2}[n_b-(1-x)n].
\end{equation}
Then, since $m_a=m_b$ this relation gives $n_a=n_b$, as it should be. It is found numerically that $x=0$ and $n=n_a=n_b$, favoring the BCS state with
energy
\begin{equation}
\label{k6}
K^{\rm MIX}(m_a=m_b)= \frac{(6\pi^2  n_a)^{5/3}}{10\pi^2 m_a}-
\frac{m_a \Delta_0^2(n_a) (6\pi^2 n_a)^{1/3}}{4\pi^2}.
\end{equation}
\newline
{\bf II:}~$m_b \gg m_a$. Plugging $n_a$ from Eq. (\ref{k5}) in Eq. (\ref{k4}) we find, in the limit $m_b \gg m_a$, that

\begin{align}
\label{k7}
K^{\rm MIX}(m_b \gg m_a)&=\frac{1}{x^{2/3}} \frac{(6 \pi^2 )^{5/3}}{20 \pi^2 m_b} [n_b -(1-x)n]^{5/3} \\
&+(1-x) \left[ \frac{(6\pi^2  n)^{5/3}}{20\pi^2 M}-
\frac{M\Delta_0^2(n) (6\pi^2 n)^{1/3}}{2\pi^2} \right].
\end{align}
This equation is minimized when $x=1$, giving
\begin{equation}
\label{k8}
K^{\rm MIX}(m_b \gg m_a)=\frac{(6 \pi^2 n_b)^{5/3}}{20 \pi^2 m_b}.
\end{equation}
This is exactly the energy of the normal state with only $b$ particles.
It is worth to mention that, in a system of fixed overall density with intermediate values of the particles masses
(different from these two limits we have considered), again the Sarma phase can not be the ground state.
Since for $\delta \tilde \mu^{\rm N}=\delta \mu^{\rm BCS}=0$ the condition for its existence (Eq. (\ref{eq27})) is not satisfied.

Then we have seen that the Sarma phase is always unstable against the mixed phase in all situations we have considered.

\section{Conclusions}
We have studied cold fermionic gases composed by two particle species whose Fermi surfaces or densities do not match. We found the ground state of such
a systems in three physically different situations, namely fixed chemical potentials, fixed number densities and fixed overall density.
In all cases the extreme limits are the normal and the BCS states, depending on the asymmetry in the particle densities or masses.

For fixed (and different) chemical potentials the system is found to be in the BCS state until a point where the asymmetry between the
Fermi surfaces is too big so there is a first order phase transition to the normal phase. We have investigated the Sarma phase, recently proposed as the ground state of asymmetrical fermion systems, employing canonical transformation, in mean-field approximation. For fixed chemical potentials, we found that the Sarma gap corresponds to an {\it extremum} of the free-energy ${\rm F}^S$. It never represents its minimum.

In the situation where the fixed parameter is the number density, we have found that the ground
state is a (inhomogeneous) mixed phase in real space, formed by islands of asymmetric normal state immersed in a sea of the symmetric BCS state. Again, if the particles $a$ and $b$ have the same density, the system is entirely in the BCS state, whereas when the asymmetry is above some limiting value, the system is found to be in the normal state. We also showed that the two components of the mixed phase state are in thermodynamical equilibrium in complete agreement with the Gibbs conditions.

The third case considered is the one with fixed overall number density, relevant for physics of high density
quark matter. The two pertinent limits for the masses of the species $a$ and $b$ in this case have been considered and we have obtained the favored ground state in these situations.

We have suggested in Ref. \cite{PRL1} that the mixed phase could be used for detecting superfluidity in atomic traps. An optical atomic trap should be
prepared having an density asymmetry in the two species of fermionic atoms. This could be done in current experiments \cite{Stoof}, by loading the trap in
a way that one of the two hyperfine states of lithium atoms (for instance) is more populated than the other.
So the islands of normal matter in a sea of BCS phase would become apparent and could be imaged. Experiments with different densities would have different segregations of the normal phase that could be located, for instance, in a corner
of the apparatus, leading to a estimative of the ``size'' of the gap, which is an up-to-date open problem.

Since we do not know how big the surface tension effects in finite systems could be, the plans for the future include the study of the magnitude of the interface energy between the superfluid and normal phases and also the investigation of the ground state of (asymmetric) neutral quark matter.

We hope that our work will stimulate new experiments with this interesting asymmetrical fermion superfluids.

\label{conc}

\section*{Acknowledgments}
The author thanks P. Bedaque for enlightening discussions, for stimulating my interest in the subject and for a critical reading of the manuscript.
I also thank G. Rupak for helpful discussions, J. Randrup for useful conversations and suggestions and L. Grandchamp for help in the figures. This work was partially supported by CAPES/Brazil.

\section{Appendix: Exact expressions and limits for the free-energy and energy in the Sarma phase}
\label{ape}
The analytical expressions for the free-energy and the energy in the Sarma phase (at the minimum), can be written as

\begin{align}
\label{eq32}
{\rm F}^{\rm S}&=-\frac{\Delta^2}{g}+\int_{\substack{k<k_1 \\ k>k_2}} \frac{d^3k}{(2 \pi)^3} ({\varepsilon}_k^+ - E)+
\int_{k_1}^{k_2} \frac{d^3k}{(2 \pi)^3} {\varepsilon}_k^b=\\
\nonumber
&-\frac{k_F^5}{30 \pi^2 M}-\frac{M k_F}{2 \pi^2} \Delta^2 -\int_{k_1}^{k_2} \frac{k^2 dk}{2 \pi^2} \left({\varepsilon}_k^+ -
\frac{ {{\varepsilon}_k^+}^2}{ \sqrt{ {{\varepsilon}_k^+}^2 +\Delta^2} } \right)+{\varepsilon}^b,
\end{align}
with ${\varepsilon}^b$ given by
\begin{equation}
\label{eq32-1}
{\varepsilon}^b=\int_{k_1}^{k_2} \frac{d^3k}{(2 \pi)^3} {\varepsilon}_k^b=
\frac{1}{4 \pi^2 m_b} \left[\frac{1}{5}(k_2^5-k_1^5)-\frac{{P_F^b}^2}{3} (k_2^3-k_1^3)\right].
\end{equation}
The limit $\Delta \to 0$ results in $k_1=P_F^a$, $k_2=P_F^b$ and Eq. (\ref{eq32}) yields the (expected) normal free-energy
\begin{align}
\label{eq32-2}
{\rm F}^{\rm N}(\mu_a,\mu_b)&={\rm F}^{\rm S}(\Delta=0)=-\frac{1}{30 \pi^2}\left[\frac{{P_F^a}^5}{m_a}+\frac{{P_F^b}^5}{m_b}\right]\\
\nonumber
&=-\frac{1}{30 \pi^2}\left[\frac{{(2 m_a \mu_a)}^{5/2}}{m_a}+\frac{{(2 m_b \mu_b)}^{5/2}}{m_b}\right].
\end{align}
When the Sarma gap has its maximum value, $k_1=k_2$ and the Sarma free-energy reduces to the BCS free-energy (Eq. (\ref{eq9}))

\begin{equation}
\label{eq32-3}
{\rm F}^{\rm BCS}(\mu)=-\frac{k_F^5}{30 \pi^2 M}-\frac{M k_F}{2 \pi^2} \Delta_0^2.
\end{equation}
The analytic expression for the energy in the Sarma phase is
\begin{align}
\label{eq32-4}
{\rm E}^{\rm S} &= {\rm F}^{\rm S} + \sum_{\alpha} \mu_{\alpha} n_{\alpha}=\frac{k_F^5}{20 \pi^2 M}+
\frac{k_F^3 \Delta^2}{8 \pi^2 \mu} \left(3 + \frac{\pi}{ |a| k_F} \right)\\
\nonumber
 &-\int_{k_1}^{k_2} \frac{k^2 dk}{2 \pi^2} \frac{k^2}{4 M} \left(1- \frac{{\varepsilon}_k^+}{E} \right) + \int_{k_1}^{k_2} \frac{k^2 dk}{2 \pi^2} \frac{k^2}{2 m_b}.
\end{align}
Again, with the limit $\Delta \to 0$ it is very easy to get the normal energy,
\begin{equation}
\label{eq32-5}
{\rm E}^{\rm N}(n_a,n_b)={\rm E}^{\rm S}(\Delta=0)=\frac{1}{20 \pi^2}\left[\frac{{P_F^a}^5}{m_a}+\frac{{P_F^b}^5}{m_b}\right]=
\frac{(6 \pi^2)^{5/3}}{20 \pi^2}\left[\frac{{n_a}^{5/3}}{m_a}+\frac{{n_b}^{5/3}}{m_b}\right].
\end{equation}
The BCS energy is recovered in the limit $\Delta \to \Delta_0$. In this case the last two terms in Eq. (\ref{eq32-4}) vanish and we obtain

\begin{equation}
\label{eq32-6}
{\rm E}^{\rm BCS}(\mu)= {\rm E}^{\rm S}(\Delta=\Delta_0)=\frac{k_F^5}{20 \pi^2 M}+
\frac{k_F^3 \Delta_0^2}{8 \pi^2 \mu} \left(3 + \frac{\pi}{ |a| k_F} \right).
\end{equation}




\begin{thebibliography}{99}

\bibitem{Marco} B. DeMarco and D.S. Jin, Science {\bf 285}, 1703 (1999).

\bibitem{Truscott} A.G. Truscott {\it et al.}, science {\bf 291}, 2570 (2001).

\bibitem{Schreck} F. Schrek {\it et al.}, Phys. Rev. Lett. {\bf 87}, 080403 (2001).

\bibitem{Granade} S.R. Granade {\it et al.}, Phys. Rev. Lett. {\bf 88}, 120405 (2002).

\bibitem{Hadzibabic} Z. Hadzibabic {\it et al.}, Phys. Rev. Lett. {\bf 88}, 160401 (2002).

\bibitem{Roati} G. Roati {\it et al.}, Phys. Rev. Lett. {\bf 89}, 150403 (2002).

\bibitem{Thomas} K.~M.~O`Hara {\it et al.}, Science {\bf 298}, 2179 (2002).

\bibitem{pethick_book}
C.~J.~Pethick and H.~Smith, {\it Bose-Einstein Condensation in Dilute Gases} (Cambridge University Press, Cambridge 2001).

\bibitem{krishna_review}
K.~Rajagopal and F.~Wilczek,
\newblock hep-ph/0011333.

\bibitem{Schafer:2003vz}
T.~Schafer,
\newblock hep-ph/0304281.

\bibitem{Rischke:2003mt}
D.~H. Rischke,
\newblock nucl-th/0305030.

\bibitem{BCS} J. Bardeen, L.N. Cooper, and J. R. Schrieffer, Phys. Rev. {\bf 108}, 1175 (1957).

\bibitem{Sarma} G. Sarma, Phys. Chem. Solid {\bf 24}, 1029 (1963).

\bibitem{Liu1} W. Vincent Liu and Frank Wilczek, Phys. Rev. Lett. {\bf 90}, 047002 (2003), cond-mat/0208052.

\bibitem{Liu2}
W.~V. Liu and F.~Wilczek,
\newblock cond-mat/0304632.

\bibitem{PRL1} P. Bedaque, H. Caldas, and G. Rupak, Phys. Rev. Lett. {\bf 91}, 247002 (2003), cond-mat/0306694.

\bibitem{larkin} A.I. Larkin and Yu. N. Ovchinnikov, Sov. Phys. JETP {\bf 20} 762 (1965);
P. Fulde and R.A. Ferrel,
\newblock Phys. Rev. {\bf 135}, A550 (1964).

\bibitem{Alford:2001dt}
M.~G. Alford,
\newblock Ann. Rev. Nucl. Part. Sci. {\bf 51}, 131 (2001), hep-ph/0102047.

\bibitem{QCD-2} J. Bowers and K. Rajagopal, hep-ph/0204079.

\bibitem{Alford} M. Alford, K. Rajagopal and F. Wilkzek, Phys. Lett. B {\bf 422}, 247 (1998), hep-ph/9711395.

\bibitem{Igor1} I. Shovkovy and M. Huang, Phys. Lett. B {\bf 564}, 205 (2003), hep-ph/0302142.

\bibitem{Igor2} M. Huang and I. Shovkovy, Nucl. Phys. A {\bf 729}, 835 (2003), hep-ph/0311155.

\bibitem{Mark} M. Alford, C. Kouvaris and K. Rajagopal, hep-ph/0311286.

\bibitem{Tinkham} M. Tinkham, {\it Introduction to Superconductivity} (MacGraw-Hill, New York 1996).

\bibitem{Pape}  T. Papenbrock and G.F. Bertsch, Phys. Rev. {\bf C 59}, 2052 (1999), nucl-th/9811077.

\bibitem{Wu} S.-T. Wu and S.~Yip,
\newblock Phys. Rev. {\bf A67}, 053603 (2003), cond-mat/0303185.

\bibitem{Elena} Elena Gubankova, W. Vincent Liu and Frank Wilczek, Phys. Rev. Lett., {\bf 91}, 032001 (2003), hep-ph/0304016.

\bibitem{Norman1} N.~K. Glendenning,
\newblock Phys. Rev. {\bf D46}, 1274 (1992).

\bibitem{Sedrakian} A. Sedrakian, Phys. Rev. {\bf C 63}, 025801 (2001), nucl-th/0008052.

\bibitem{Paulo} P. Bedaque, Nuc. Phys. {\bf A 697}, 569 (2002), hep-ph/9910247.

\bibitem{Buballa} F. Neuman, M. Buballa and M. Oertel, Nuc. Phys. {\bf A 714}, 481 (2003), hep-ph/0210078.

\bibitem{Stoof} H.T.C. Stoof {\it et al.}, Phys. Rev. Lett. {\bf 76}, 10 (1996), cond-mat/9508079.





\end{thebibliography}
\end{document}